\documentclass[aps,prd,twocolumn,showpacs,nofootinbib]{revtex4-1}

\usepackage{amssymb} \usepackage{amsmath} \usepackage{graphicx}
\usepackage{epsfig,latexsym}
\usepackage{mathtools}

\RequirePackage{xspace} \allowdisplaybreaks

\begin{document}

\def\bef{\begin{figure}}
\def\eef{\end{figure}}

\newcommand{\nl}{\nonumber\\}

\newcommand{\ans}{ansatz }
\newcommand{\be}[1]{\begin{equation}\label{#1}}
\newcommand{\beq}{\begin{equation}}
\newcommand{\ee}{\end{equation}}
\newcommand{\beqn}[1]{\begin{eqnarray}\label{#1}}
\newcommand{\eeqn}{\end{eqnarray}}
\newcommand{\bd}{\begin{displaymath}}
\newcommand{\ed}{\end{displaymath}}
\newcommand{\mat}[4]{\left(\begin{array}{cc}{#1}&{#2}\\{#3}&{#4}
\end{array}\right)}
\newcommand{\matr}[9]{\left(\begin{array}{ccc}{#1}&{#2}&{#3}\\
{#4}&{#5}&{#6}\\{#7}&{#8}&{#9}\end{array}\right)}
\newcommand{\matrr}[6]{\left(\begin{array}{cc}{#1}&{#2}\\
{#3}&{#4}\\{#5}&{#6}\end{array}\right)}
\newcommand{\cvb}[3]{#1^{#2}_{#3}}
\def\lsim{\raise0.3ex\hbox{$\;<$\kern-0.75em\raise-1.1ex
e\hbox{$\sim\;$}}}
\def\gsim{\raise0.3ex\hbox{$\;>$\kern-0.75em\raise-1.1ex
\hbox{$\sim\;$}}}
\def\abs#1{\left| #1\right|}
\def\simlt{\mathrel{\lower2.5pt\vbox{\lineskip=0pt\baselineskip=0pt
           \hbox{$<$}\hbox{$\sim$}}}}
\def\simgt{\mathrel{\lower2.5pt\vbox{\lineskip=0pt\baselineskip=0pt
           \hbox{$>$}\hbox{$\sim$}}}}
\def\unity{{\hbox{1\kern-.8mm l}}}
\newcommand{\eps}{\varepsilon}
\def\ep{\epsilon}
\def\ga{\gamma}
\def\Ga{\Gamma}
\def\om{\omega}
\def\omp{{\omega^\prime}}
\def\Om{\Omega}
\def\la{\lambda}
\def\La{\Lambda}
\def\al{\alpha}
\newcommand{\ov}{\overline}
\renewcommand{\to}{\rightarrow}
\renewcommand{\vec}[1]{\mathbf{#1}}
\newcommand{\vect}[1]{\mbox{\boldmath$#1$}}
\def\tm{{\widetilde{m}}}
\def\mcirc{{\stackrel{o}{m}}}
\newcommand{\Dm}{\Delta m}
\newcommand{\dm}{\varepsilon}
\newcommand{\tanb}{\tan\beta}
\newcommand{\nbar}{\tilde{n}}
\newcommand\PM[1]{\begin{pmatrix}#1\end{pmatrix}}
\newcommand{\up}{\uparrow}
\newcommand{\down}{\downarrow}
\def\omE{\omega_{\rm Ter}}

%

\newcommand{\Dsusy}{{susy \hspace{-9.4pt} \slash}\;}
\newcommand{\DCP}{{CP \hspace{-7.4pt} \slash}\;}
\newcommand{\mc}{\mathcal}
\newcommand{\gr}{\mathbf}
\renewcommand{\to}{\rightarrow}
\newcommand{\gtc}{\mathfrak}
\newcommand{\wh}{\widehat}
\newcommand{\br}{\langle}
\newcommand{\kt}{\rangle}


\def\lsim{\mathrel{\mathop  {\hbox{\lower0.5ex\hbox{$\sim$}
\kern-0.8em\lower-0.7ex\hbox{$<$}}}}}
\def\gsim{\mathrel{\mathop  {\hbox{\lower0.5ex\hbox{$\sim$}
\kern-0.8em\lower-0.7ex\hbox{$>$}}}}}

\def\nn{\\  \nonumber}
\def\de{\partial}
\def\brf{{\mathbf f}}
\def\bbf{\bar{\bf f}}
\def\bF{{\bf F}}
\def\bbF{\bar{\bf F}}
\def\bA{{\mathbf A}}
\def\bB{{\mathbf B}}
\def\bG{{\mathbf G}}
\def\bI{{\mathbf I}}
\def\bM{{\mathbf M}}
\def\bY{{\mathbf Y}}
\def\bX{{\mathbf X}}
\def\bS{{\mathbf S}}
\def\bb{{\mathbf b}}
\def\bh{{\mathbf h}}
\def\bg{{\mathbf g}}
\def\bla{{\mathbf \la}}
\def\bmu{\mathbf m }
\def\by{{\mathbf y}}
\def\bmu{\mbox{\boldmath $\mu$} }
\def\bsig{\mbox{\boldmath $\sigma$} }
\def\bunity{{\mathbf 1}}
\def\cA{{\cal A}}
\def\cB{{\cal B}}
\def\cC{{\cal C}}
\def\cD{{\cal D}}
\def\cF{{\cal F}}
\def\cG{{\cal G}}
\def\cH{{\cal H}}
\def\cI{{\cal I}}
\def\cL{{\cal L}}
\def\cN{{\cal N}}
\def\cM{{\cal M}}
\def\cO{{\cal O}}
\def\cR{{\cal R}}
\def\cS{{\cal S}}
\def\cT{{\cal T}}
\def\eV{{\rm eV}}

%

\title{Chaotic instantons in scalar field theory}

\author{Andrea Addazi$^1$}\email{andrea.addazi@infn.lngs.it}
\affiliation{$^1$ Dipartimento di Fisica,
 Universit\`a di L'Aquila, 67010 Coppito AQ and
LNGS, Laboratori Nazionali del Gran Sasso, 67010 Assergi AQ, Italy}

\begin{abstract}

We consider a new class of instantons in context of quantum field theory of a scalar field
coupled with a chaotic background source field. 
We show how the instanton associated to the 
quantum tunneling from a metastable false to the true vacuum 
will be corrected by an exponential enhancement factor. 
Possible implications are discussed. 

\end{abstract}

\maketitle
\section{Introduction}

As is known, an instanton describes a quantum process 
of tunneling through a barrier. In non-relativistic quantum mechanics,
the barrier is usually localized in the space, $V({\bf r})$. 
On the other hand, in quantum field theory, the tunneling process is more subtle
because it happens in the field internal space: 
a spin 0 particle, with quantum field operator $\varphi$, can tunnel from 
a false vacua 
to the global minimum of its interaction potential $V(\varphi)$. 
This implies that not all gauge theories 
have a stable vacuum. In cosmology, a quantum tunneling can
lead to the nucleation of a Bubble or a new Baby Universe
from the Mother Universe \cite{Coleman:1977py,Callan:1977pt,Coleman:1980aw}. 
As known, the tunneling probability 
can be enhanced in presence of a thermal bath.
The most relevant theoretical example is in context of the 
baryogenesis in the Early Universe, 
where $B+L$ violating transitions ($B-L$ preserving), mediated by
Standard Model sphalerons, are enhanced by 
thermal background fluctuations while 
strongly suppressed in laboratory
 \cite{Kuzmin:1985mm,Kuzmin:1987wn}.
The possibility of a long-lived metastable 
vacuum was also largely studied in many models with dynamical supersymmetry breaking
\cite{Intriligator:2006dd}. 

On the other hand, a simple class of  
 chaotic instantons 
 in non-relativistic $1+1$ quantum mechanics 
 was studied in Refs.
\cite{Kuvshinov:2003yy,Kuvshinov:2008sz}.
In particular, the wave function of a particle can be chaotized by a series of kicks
inside a potential. The chaotic fluctuations can assist the 
quantum tunneling through the wall. 
So that, it is conceivable that a subtler analogous process can be 
considered in context of quantum field theories. 
Previous studies and definitions of 
chaos in quantum field theory can be found in Refs.
\cite{Kuvshinov:2002ww,Kuvshinov:2003kc,Kuvshinov:2002xq}.
However, chaos instantons in quantum field theory were never suggested and calculated in literature.
On the other hand, studies of possible crossroads among chaos theory and field theories 
can deserve some new surprising effects. 
For instance, in our previous papers, we have shown that Sinai billiards of singular geometries  
 can scatter and chaotize quantum wave functions \cite{Addazi:2015gna,Addazi:2015hpa,Addazi:2015bee,Addazi:2016cad}.
 A new quantum decoherence effect induced by the non-trivial space-time topology emerged 
 with possible implications in black hole physics and cosmic strings web.
\begin{figure}[htb] \label{CHAOTICI}
\begin{center}
\caption{A scalar field $\varphi$ with a generic potential having two minima $\varphi_{\pm}$. If $V(\varphi_{-})>V(\varphi_{+})$ the particle can 
quantum tunnel through the barrier from the false to the global minimum. However, the presence of a turbulent background source (drawn as a zigzag line)
coupled with the scalar field can exponentially enhance the tunneling probability. 
The process is mediated by a new instanton solution, dub chaotic instanton. }
\includegraphics[scale=0.08]{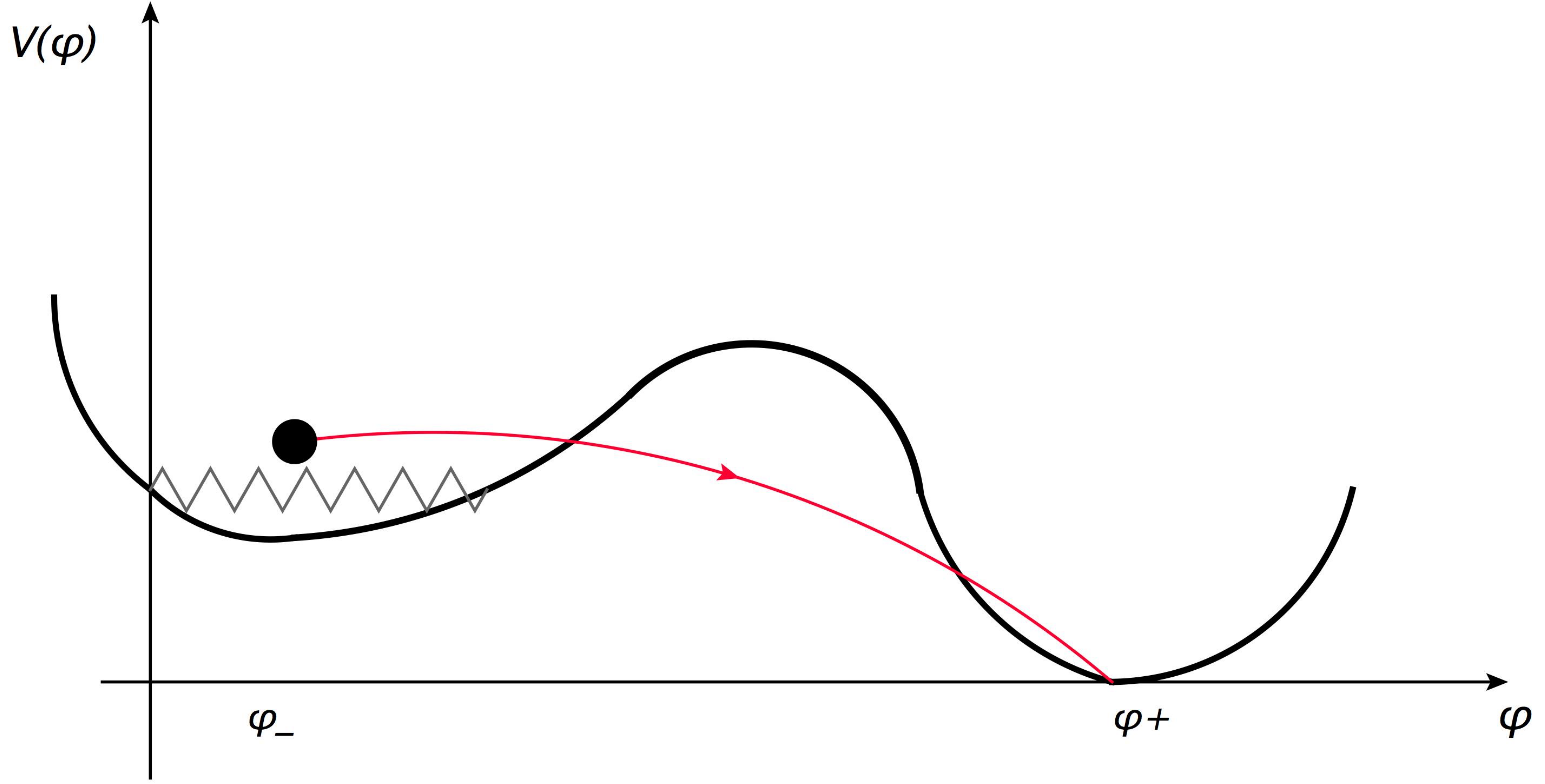}  
\end{center}
\vspace{-1mm}
\end{figure}

In this paper, we will show how a new class of chaos assisted instantons can be found in 
quantum field theory of a scalar coupled to a chaotic background source field. 
We will show that the quantum tunneling probability is exponentially 
enhanced by chaotic fluctuations. Our calculations will be considered in semiclassical regime, 
treating the interactions of the scalar field with the background source as 
a perturbation.

\section{Chaotic Instanton Solution}

Let us consider the 
\be{per}
V(\varphi)=V_{0}+aV_{per}
\ee
\be{Vper}
V_{pert}=  \, \varphi \, \Sigma= \, \varphi \sigma_{0}\sum \cos(2\pi n \hat{k}_{\mu}x^{\mu}/X)
\ee
where we define $\Sigma$ {\it  chaotic background source field}, decomposed as an Fourier series of space-time plane waves 
with $X$ a characteristic scales with dimensions $[X]=M^{-1}$, $[\sigma_{0}]=M^{3}$ and $V_{0}$ is a generic double-well potential shown in Fig.1, 
$a$ is a small dimensionless coupling constant with dimension $[a]=M^{0}$, while $[\sigma_{0}]=M^{3}$. 
This interaction describes a particle in the double-well potential 
coupled with a chaotic background field. 
With $V_{pert}=0$, the quantum tunneling process is described 
by the usual Coleman-De Luccia instanton \cite{Coleman:1977py,Callan:1977pt,Coleman:1980aw}. 
However, the perturbation forms a stochastic energy layer,
and the tunneling probability is described by 
chaotic instantons. Assuming that the stochastic corrections 
can be treated as perturbation, the chaotic instanton has a form
\be{Xchaos}
\varphi=\varphi_{0}+\delta \varphi
\ee
and they are solutions of the Euclidean equations of motion
\be{phi}
-\Box_{E} \varphi+V_{0}'(\varphi)+a\Sigma=0
\ee
with 
\be{phi}
-\Box_{E} \varphi_{0}+V_{0}'(\varphi_{0})=0
\ee
leading to the 
\be{comb}
-\Box_{E} \delta \varphi+ \delta V'_{0}+   a\Sigma=0
\ee
$$\delta V'_{0}=V'_{0}(\varphi_{0}+\delta \varphi)-V_{0}(\varphi_{0})$$
$$=3\lambda\varphi^{3}\delta \varphi-2m^{2}\varphi \delta \varphi+O(\delta \varphi)$$
On the other hand, $\Sigma$ can be rewritten in term of a Dirac comb in space-time \cite{DiracComb}
\be{DiracComb}
\sigma_{0}\sum \cos( 2\pi n \hat{k}_{\mu}x^{\mu}/X)=\sigma_{0}\mathcal{N}'\sum_{n'=-\infty}^{+\infty}\delta(x-n' X)
\ee
where $\mathcal{N}'=V_{D}$ and $V_{D}=X^{D}$.

Incidently, Let us note that setting $V'-m^{2}=0$, the equation can be solved with the method of Green's functions
and the superposition principle:
\be{phi}
\varphi(y_{E})=\sum_{n'}\varphi_{n'}(y_{E})
\ee
\be{super}
\varphi_{n'}(y_{E})=\int d^{D}x_{E}G_{n'}(y_{E},x_{E}) \delta(y_{E}-x_{E}-n' X_{E})
\ee
\be{Box}
(-\Box_{E} -m^{2})G_{n'}(y_{E},x_{E})=-4\pi \delta(y_{E}-x_{E}-n X_{E})
\ee
The retarded Green's function is 
\be{Gret}
G^{>}_{n'}(y_{E},x_{E})=\theta(y_{E}^{0}-x_{E}^{0}-n' X_{E})[\delta(z_{n'})
\ee
$$-\frac{m}{\sqrt{-2z_{n'}}}J_{1}(m\sqrt{-2z_{n'}})]$$
while the advanced Green's function is 
\be{Gadv}
G^{<}_{n'}(y_{E},x_{E})=\theta(-y_{E}^{0}-x_{E}^{0}-n' X_{E})[\delta(z_{n'})
\ee
$$-\frac{m}{\sqrt{-2z_{n'}}}J_{1}(m\sqrt{-2z_{n'}})\theta(-z_{n'})]$$
with 
$$z_{n'}=\frac{1}{2}\eta^{ab}_{E}(y_{E}^{a}-x_{E}^{a}-n X_{E}^{a})(y_{E}^{b}-x_{E}^{b}-n' X_{E}^{b})$$
and $J_{1}$ the first order Bessel function. 
So that, $\varphi$ is a noisy superposition of harmonics $\varphi_{n'}$. 

In $D=4$, the euclidean action is 
\be{SE}
S_{E}=4\pi^{2}\int_{0}^{\infty}d\rho\rho^{3}\left[\frac{1}{2}\varphi'^{2}+V_{0}(\varphi)+a\varphi \Sigma(\rho)\right]
\ee
with a resulting normalized factor 
\be{norm}
K=\frac{\bar{S}_{E}^{2}}{4\pi^{2}}\sqrt{\frac{-\Box_{E}+V''(\varphi_{0})}
{-\Box_{E}+V''(\bar{\varphi})}}
\ee
while the formal tunneling rate is 
\be{GammaV}
\frac{\Gamma}{V}=Ke^{-\bar{S}_{E}}
\ee
which can be rewritten as 
\be{Gamma00}
\frac{\Gamma}{V}=\frac{\bar{S}_{E}^{2}}{4\pi^{2}}e^{-(\tilde{\Gamma}[\bar{\varphi}]-\tilde{\Gamma}[\varphi_{0}])}
\ee
where 
\be{GammaE}
\tilde{\Gamma}_{E}[\varphi]=S_{E}[\varphi]-\frac{1}{2}{\rm Re\, tr} \log\left(-\Box_{E}+V''(\varphi) \right)
\ee

and the equation of motion is 
\be{phidue}
\varphi''+\frac{3}{\rho}\varphi'=\frac{dV_{0}}{d\varphi}+a\Sigma
\ee
in which, as done by Coleman and De Luccia , we neglect the term $\frac{3}{\rho}\varphi'$ in Eq.\ref{phidue}, 
i.e. physically motivated by a thin-wall approximation: 
\be{phidue}
\varphi''\simeq \frac{dV_{0}}{d\varphi}+a\Sigma
\ee


Let us note that in WKB approximation $\lambda_{\varphi}^{Compton}>>X$, we can perform 
a continuos spectrum approximation, and the background source is reduced to a constant
$\Sigma=\sigma_{0}\,\mathcal{N}'\sum_{n'} \delta \rightarrow \sigma_{0}$
where $\sum_{n'}\rightarrow \int dn'(...)$
and $\int dn' \delta=1/\mathcal{N}'$. 
This term becomes formally analogous to a driven constant force in an anharmonic oscillator. 
This corresponds to an energy density  scale $a\sigma_{0} \delta \epsilon_{C}$, 
where $\delta \epsilon_{C}$ is the chaotic energy layer, 
measured by the bottom of the false minima. 
The chaotic layer is 
$\delta \epsilon_{C}=2\pi r_{0}^{-1}$, where $r_{0}$ is the radial constant 
related to $X$ as
$X^{4}=\pi^{2}r_{0}^{4}/2$. 
Now, let us suppose that the background $\Sigma$ is localized in space-time volume with radius $R\leq \bar{\rho}$,
i.e. $R=f \bar{\rho}$  with $0\leq f\leq 1$. 
Physically, it will positively contribute to the formation of a bubble in this space-time radius. 
Outside the wall $\phi=\phi_{+}$ the euclidean action is 
\be{BA}
\bar{S}_{E,(\phi=\phi_{+})}=0
\ee
where inside the wall $\phi=\phi_{-}$
\be{BA2}
\bar{S}_{E,(\phi=\phi_{-})}=-\frac{\pi^{2}}{2}\bar{\rho}^{4}\epsilon+\pi^{3}f^{4}\bar{\rho}^{4}a\sigma_{0}r_{0}^{-1}
\ee
where $$\epsilon=V_{0}(\varphi_{+})-V_{0}(\varphi_{-})$$
Within the wall:
\be{within}
\bar{S}_{E,(\phi_{-}<\phi<\phi_{+})}= 2\pi^{2} \bar{\rho}^{3}\int d\rho \left[V_{0}(\phi)-V_{0}(\phi_{+})\right]
\ee
So that, the total $\bar{S}_{E}$ is 
\be{total}
\bar{S}_{E}=-\frac{\pi^{2}}{2}\bar{\rho}^{4}\epsilon+2\pi^{2}\bar{\rho}^{3}s_{1}+\pi^{3}f^{4}\bar{\rho}^{4}a\sigma_{0}r_{0}^{-1}
\ee
which is stationary if 
$$\bar{\rho}=3s_{1}/(\epsilon-2\pi af^{4}\sigma_{0}r_{0}^{-1})$$
The expression is well defined for $2\pi af^{4}\sigma_{0}r_{0}^{-1}<\epsilon$.
Finally, the tunneling amplitude is 
\be{A}
\frac{1}{V}\Gamma\simeq Ke^{S_{0}+\pi^{3}f^{4}\bar{\rho}^{4}a\sigma_{0}r_{0}^{-1}}
\ee
So that, as we can see, an exponential enhancement factor with respect to the standard instanton.

\section{Conclusions and remarks}

In this paper, we have considered the problem of a scalar field 
in a false vacuum coupled to a chaotic background source field. 
The transition amplitude gets an extra exponential correction with respect to the 
result obtained by Coleman {\it et al.} \cite{Coleman:1977py,Callan:1977pt,Coleman:1980aw}.
This is the first example of chaos assisted instantonic solution
in quantum field theory. 
However, it is conceivable that the solution suggested in this paper 
is only one particular example in a large class of possible solutions of various quantum field theories. 
For instance, it could be interesting to study possible chaos instantons in 
gauge theories. Perhaps,  chaos gauge instantons could provide important insights in 
QCD asymptotic freedom, confinement and nuclear physics decays. 
On the other hand, in context of early Universe, chaos assisted enhancements can 
induce new first order phase transitions. 
As mentioned above, we have not include gravity in our considerations.
Coleman and De Luccia demonstrated that the inclusion of gravity 
implies that the transition from a false vacuum generates a 
Bubble \cite{Coleman:1980aw}. 
Now, let us comment on possible 
existence of chaotic solitons. 
In fact a general correspondence
among instantons in gauge theories with solitons in higher dimension
is known in literature. 
 G.~Dvali, H.~B.~Nielsen and N.~Tetradis have shown a correspondence 
among the t'Hooft-Polyakov monopole and 
a gauge instanton solution in the lower dimensional theory 
localized on the domain walls \cite{Dvali:2007nm}. Other examples and generalizations of the 
instanton/monopole correspondence were studied in literature  
  \cite{Auzzi:2008zd,Grigorio:2008gd,Flassig:2011qh,Addazi:2016yre}.
  So that, if the scalar field considered above is an Higgs field, 
  it can source the presence of monopoles, vortices or cosmic strings 
  in context of GUT theories. For example monopoles, kinks, vortices and domain walls
  in false vacua were studied in 
  \cite{Kumar:2010mv,Haberichter:2015xga,Dupuis:2015fza,Lee:2013ega}
: topological solitons can tunnel with a finite probability 
  mediated by a gauge instantons (of a lower dimensional gauge theory). 
So that, we suggest that in presence of a chaotic source, 
new solitons could correspond to chaotic instantons. 
These considerations could be extended in context of open string theories, 
where instantons correspond to Euclidean D-branes wrapping the internal Calabi-Yau compactification
(see \cite{Tong:2005un,Bianchi:2009ij,rev1,rev2,rev3} for general reviews on these aspects). 
Finally, it is possible that in presence of a primordial chaotic background 
electroweak gauge instantons or sphalerons get an extra enhancement factor 
higher then the thermally induced one. 
On the other hand, the chaotic enhancement effects could 
induce strong violations of $B-L$,
if associated exotic instantons 
were assisted by a chaotic background
  \cite{Addazi:2014ila,Addazi:2015ewa,Addazi:2015goa}.

\vspace{0.3cm}

{\large \bf Acknowledgments}
\vspace{3mm}

I also would like to thank M. Bianchi and G. Dvali 
for useful discussions and remarks on these subjects. 
My work was supported in part by the MIUR research grant Theoretical Astroparticle Physics PRIN 2012CP-PYP7 and by SdC Progetto speciale Multiasse La Societ\'a della Conoscenza in Abruzzo PO FSE Abruzzo 2007-2013.

\end{document}